\documentclass[aps,prb,reprint,superscriptaddress]{revtex4-1}
\usepackage{amsmath}
\usepackage{amsfonts}
\usepackage{graphicx}
\usepackage{hyperref}
\usepackage{color}
\usepackage{enumerate}
\newcommand{\eeT}{\ee_{\mathrm{TOP}}}

\newcommand{\eeH}{\ee_{\mathrm{HS}}}

\newcommand{\ThetaT}{\Theta_{\mathrm{TOP}}}
\newcommand{\ThetaH}{\Theta_{\mathrm{HS}}}
\newcommand{\ThetaV}{\Theta_{\mathrm{VAC}}}
\newcommand{\ThetaR}{\Theta_{\mathrm{REC}}}
\newcommand{\GammaT}{\Gamma_{\mathrm{TOP}}}
\newcommand{\GammaH}{\Gamma_{\mathrm{HS}}}
\newcommand{\GammaV}{\Gamma_{\mathrm{VAC}}}
\newcommand{\GammaR}{\Gamma_{\mathrm{REC}}}
\newcommand{\real}[1] {\mathrm{Re}\, #1 \,}
\newcommand{\imag}[1]{\mathrm{Im}\, #1 \,}

\newcommand{\ee}{\varepsilon}

\newcommand{\abs}[1]{\left| #1 \right|}

\newcommand{\rgd}[1]{G_{I}(#1) }

\newcommand{\Hg}{H_{\mathrm{G}}}
\newcommand{\Himp}{H_{\mathrm{I}}}
\newcommand{\HtopA}{H_{\text{TOP-G}}}
\newcommand{\HvacA}{H_{\text{VAC-G}}}
\newcommand{\HrecA}{H_{\text{REC-G}}}
\newcommand{\Hhs}{H_{\text{HS-G}}}
\newcommand{\Hig}{H_{\text{I-G}}}
\newcommand{\uu}{\hat{\mathbf{u}}}

\newcommand{\kk}{\vec{k}}
\newcommand{\kp}{\vec{\kappa}}
\newcommand{\vsig}{\vec{\sigma}}

\renewcommand{\vec}[1]{\boldsymbol{\mathbf{#1}}}
\renewcommand{\exp}[1]{\mathrm{e}^{#1}}

\begin{document}

\title{Transport signatures of Kondo physics and quantum criticality in graphene with magnetic impurities.}

\author{David A.\ Ruiz-Tijerina}
\affiliation{Instituto de F\'{\i}sica, Universidade de S\~{a}o Paulo,
C.P.\ 66318, 05315--970 S\~{a}o Paulo, SP, Brazil}
\affiliation{National Graphene Institute, University of Manchester, Manchester M13 9PL, UK}
\affiliation{School of Physics and Astronomy, University of Manchester, Manchester M13 9PL, UK}

\author{Luis G.\ G.\ V.\ Dias da Silva}
\affiliation{Instituto de F\'{\i}sica, Universidade de S\~{a}o Paulo,
C.P.\ 66318, 05315--970 S\~{a}o Paulo, SP, Brazil}

\date{\today}

\begin{abstract}
Localized magnetic moments have been predicted to develop in graphene samples with vacancies or adsorbates. The interplay between such magnetic impurities and graphene's Dirac quasiparticles leads to remarkable many--body phenomena, which have so far proved elusive to experimental efforts. In this article we study the thermodynamic, spectral and transport signatures of quantum criticality and Kondo physics of a dilute ensemble of atomic impurities in graphene. We consider vacancies and adatoms that either break or preserve graphene's $C_{3v}$ and inversion symmetries. In a neutral graphene sample all cases display symmetry--dependent quantum criticality, leading to enhanced impurity scattering for asymmetric impurities, in a manner analogous to bound--state formation by nonmagnetic resonant scatterers. Kondo correlations emerge only in the presence of a back gate, with estimated Kondo temperatures well within the experimentally accessible domain for all impurity types. For symmetry--breaking impurities at charge neutrality, quantum criticality is signaled by $T^{-2}$ resistivity scaling, leading to full insulating behavior at low temperatures, while low--temperature resistivity plateaus appear both in the non--critical and Kondo regimes. By contrast, the resistivity contribution from symmetric vacancies and hollow--site adsorbates vanishes at charge neutrality and for arbitrary back gate voltages, respectively. This implies that local probing methods are required for the detection of both Kondo and quantum critical signatures in these symmetry--preserving cases.
\end{abstract}

\maketitle

\section{Introduction}\label{sec:intro}

Graphene is a two--dimensional carbon allotrope characterized by low energy excitations that behave like two--dimensional, massless Dirac fermions.\cite{semenoff_prl_1984} Many desirable electronic properties have been predicted for graphene as a result of the chiral nature of its charge carriers,\cite{castro-neto_rmp_2009} making it an ideal platform for the fabrication of novel devices. These properties, along with its outstanding mechanical characteristics, have earned graphene a place in contemporary popular culture as the ``material of the future."

Combining graphene's electronic properties with magnetism has become a major driving force for research on this material in recent years. Although ferromagnetic order is suppressed in pristine bulk samples,\cite{Peres:Phys.Rev.B:174406:2005} there are a number of ways to induce magnetic moments in graphene, ranging from adsorption of magnetic \cite{eelbo_prl_2013, donati_prl_2013,donati_prl_2014} and non--magnetic \cite{mccreary_prl_2012,Gonzalez-Herrero2016} atoms to proposals for vacancy--induced  $\pi$-magnetism.\cite{Yazyev2010,Chen:NaturePhys.:535:2011,Nair2012,Nair2013,JustPRB2014,Zhang2016}

Once magnetic moments are present in the graphene lattice, fascinating many--body effects can be expected. The coupling of a localized magnetic moment and the (real) spin of Dirac electrons in neutral graphene has long been theorized as a realization of the linear pseudogap Kondo model,\cite{WITHOFF:1835:1990,gonzalez-buxton_prb_1998} where a magnetic impurity couples to a fermionic density of states that vanishes linearly at the Fermi energy. One of the main features of this type of system is a quantum phase transition from a local--moment phase to a Kondo phase, where the atom's magnetic moment becomes screened by itinerant electrons. \cite{WITHOFF:1835:1990,gonzalez-buxton_prb_1998,fritz_vojta_kondo_in_graphene,VojtaEPL:27006:2010,uchoa_prl_2011}

Experimental evidence for the Kondo effect has emerged recently in a graphene system with atomic vacancies.\cite{Chen:NaturePhys.:535:2011} A most intriguing feature of those experiments is an observed temperature dependence of the resistivity that is compatible with the metallic Kondo effect for a vast range of gate voltages, including at the charge neutrality point. This leads us to two questions: Why are no pseudogap effects observed? And, what would the transport signatures of impurity quantum criticality be? One possible answer to the first question is that sample disorder may play an important role\cite{Miranda14} in masking the pseudogap signatures. Another possibility is that in the experiments the samples are only probed effectively away from charge neutrality, where metallic Kondo behavior predominates.\cite{cornaglia_prl_2009,kanao_jpsj_2012,po-wei-lo_prb_2014} 

In this paper we address the second question by calculating the impurity contribution to the linear resistivity of a graphene sample. In a recent work\cite{ruiz-tijerina_2016} we calculated the transport properties of graphene in the presence of a dilute ensemble of non--magnetic impurities, and found that impurity symmetry plays a determinant role in the system transport properties. Here we turn our attention to \textit{magnetic} impurities, such as transition--metal\cite{eelbo_prl_2013, donati_prl_2013,donati_prl_2014} and hydrogen\cite{mccreary_prl_2012,Gonzalez-Herrero2016} adatoms, and both symmetric and reconstructed vacancies.\cite{yazyev_prb_2007,Cazalilla12,Miranda2016,rodrigo_carbon_2016}

Using numerical renormalization--group (NRG) calculations we found that the interplay between symmetry and strong correlations leads to a rich phenomenology, including an impurity--dependent quantum phase transition (QPT) in charge neutrality related to the pseudogap Anderson model, and the emergence of Kondo physics when a back gate is applied. For top--site adsorbates and reconstructed vacancies lacking $C_{3v}$ and inversion symmetry, both regimes produce experimentally accessible signatures in the bulk resistivity profile: the QPT is signaled by a power--law divergence of the resistivity at low temperatures, when the energy of the impurity level crosses a critical value. Surprisingly, this behavior is qualitatively similar to that of the non--interacting case,\cite{ruiz-tijerina_2016} with the main effect of interactions being a renormalization of the critical point parameters. By contrast, the presence of a back gate driving the system away from charge neutrality leads to screening of the impurity spin through Kondo correlations, signaled by a zero--energy peak in the impurity density of states and a low--temperature plateau of enhanced resistivity.

In the case of highly symmetric magnetic impurities, such as unreconstructed vacancies ($C_{3v}$) and hollow--site adsorbates ($C_{3v}$ and inversion), quantum criticality is also observed in neutral graphene, described by the cubic pseudogap Anderson model. Furthermore, these impurities display robust Kondo physics for sensible values of the carrier density, with Kondo temperatures comparable to those obtained for their  less symmetric counterparts. Interestingly, we find that such impurities do not contribute to the sample resistivity, a result we had previously obtained in the non--interacting case.\cite{ruiz-tijerina_2016,ferreira_prb_2016} As a consequence, they are not easily captured in bulk transport measurements and need to be probed locally.

The remainder of this article is organized as follows: In Sec.\ \ref{sec:model} we introduce low--energy Hamiltonians for graphene coupled to four different impurity types. These models are studied in Sec.\ \ref{sec:results} using NRG calculations. We begin by describing the single--impurity ground state for different model parameters in Sections \ref{sec:QPT} and \ref{sec:SpecDens}, where we discuss an impurity--dependent QPT in charge neutrality and the onset of Kondo correlations when a back gate is applied. Then, in Sec.\ \ref{sec:resistivity} we evaluate the impurity contribution to the resistivity as a function of temperature and chemical potential, in graphene samples with mixtures of symmetric and non--symmetric adatoms or vacancies. Our conclusions are presented in Sec.\ \ref{sec:conclusions}.

\section{Model}\label{sec:model}
\begin{figure}[t]
\begin{center}
\includegraphics[width=0.95\columnwidth]{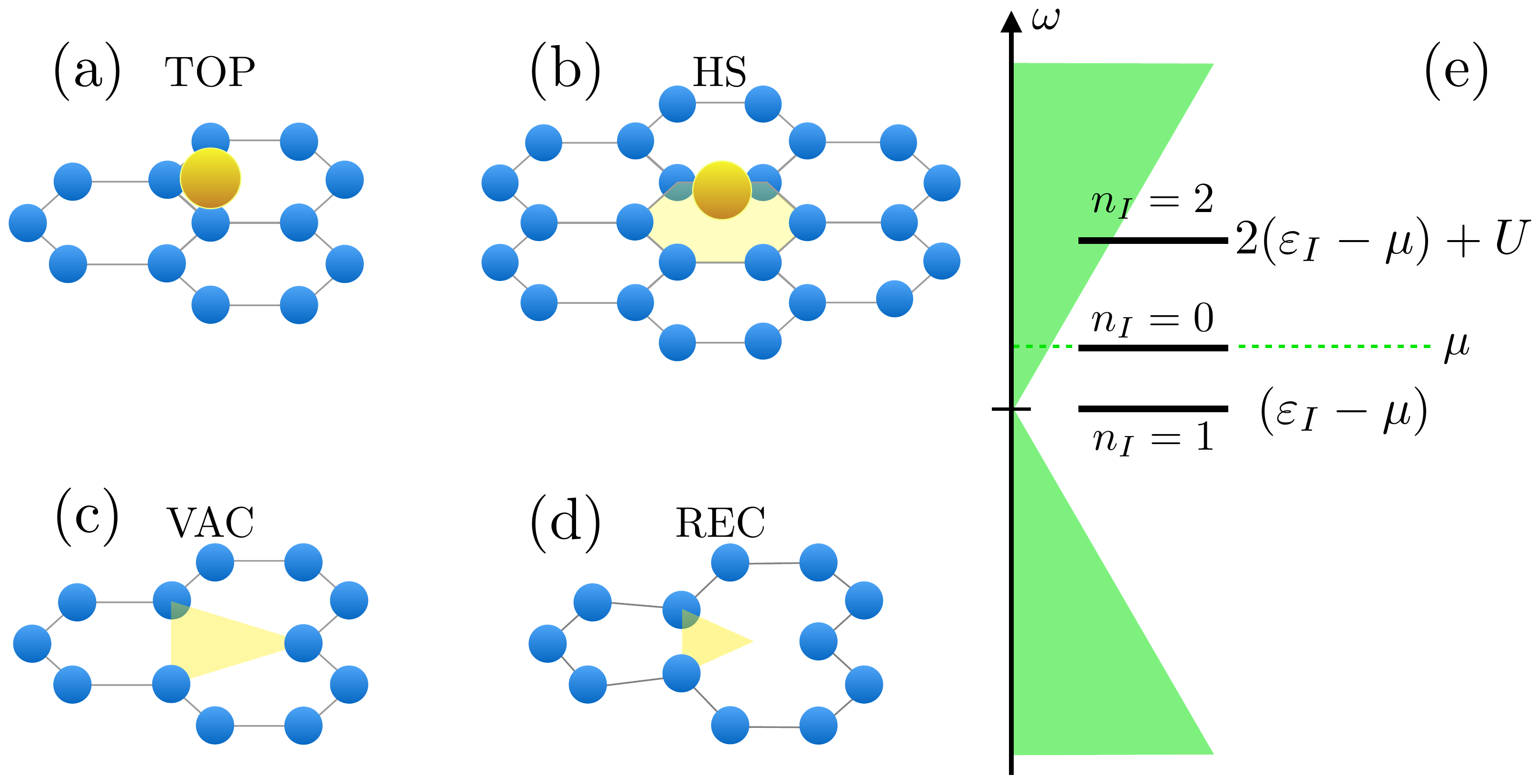}
\caption{(Color online) Atomic impurities in graphene: (a) Top--site adsorbate, (b) hollow--site adsorbate, (c) symmetric vacancy, and (d) reconstructed vacancy. (e) Graphene density of states and many--body level structure of an impurity with local energy $\varepsilon_I<0$ and Coulomb interaction $U>0$, for $\mu > 0$.}
\label{fig:impurities}
\end{center}
\end{figure}

A single magnetic impurity in a graphene sample can be modeled by the Anderson--type Hamiltonian $H = \Himp + \Hg + \Hig$, where $\Himp$ represents the impurity, $\Hg$ the graphene sheet, and $\Hig$ is a hybridization term. The interacting impurity Hamiltonian is
\begin{equation}\label{eq:Himp}
    \Himp = \sum_{s}(\ee_I-\mu)\,d_{I,s}^\dagger d_{I,s} + U\, n_{I,\uparrow}n_{I,\downarrow},
\end{equation}
where ($\ee_I - \mu$) is the impurity orbital energy measured with respect to the graphene chemical potential, $U$ is the electrostatic energy cost of double occupancy, the operators $d_{I,s}^\dagger$ ($d_{I,s}$) create (annihilate) electrons of spin projection $s$ in the impurity orbital, and $n_{I,s} = d_{I,s}^\dagger d_{I,s}$ are the corresponding number operators. The graphene sample can be described by the Hamiltonian
\begin{equation}\label{eq:Hg_ab_k}
    \Hg = -t\sum_{\kk s}\psi_s^\dagger(\kk) h(\kk) \psi_s(\kk) \,; \, h(\kk) = \begin{pmatrix} 0 & \Phi(\kk)\\\Phi^*(\kk) & 0 \end{pmatrix},
\end{equation}
with $\psi_s(\kk)=(a_s(\kk),\,b_s(\kk))^T$, $a_s(\kk)$ [$b_s(\kk)$] the annihilation operator for sublattice $A$ ($B$) electrons, and
\begin{equation}\label{eq:Phi}
    \Phi(\vec{k})\equiv\sum_{j=1}^3 \exp{ia\vec{k}\cdot\uu_j}.
\end{equation}
$a \approx 1.42\,\mathrm{\r{A}}$ is the honeycomb lattice constant, $t$ the nearest--neighbor hopping energy, and $\uu_1 = \hat{\mathbf{x}}$, $\uu_2 = -\hat{\mathbf{x}}/2 + \hat{\mathbf{y}}\sqrt{3}/2$ and $\uu_3 = -\hat{\mathbf{x}}/2 - \hat{\mathbf{y}}\sqrt{3}/2$ are nearest--neighbor vectors. Diagonalizing $\Hg$ gives two energy bands with dispersions $\epsilon_{\pm}(\vec{k}) = \pm t|\Phi(\vec{k})|$ and corresponding operators $c_s(\kk)=(c_{+,s}(\kk),\,c_{-,s}(\kk))^T$. These are related to the $\psi$-basis operators as $\psi_s(\kk) = U_{\vec{k}}c_s(\kk)$, with the similarity transformation
\begin{equation}\label{eq:unitary_2band}
    U_{\vec{k}} = \frac{1}{\sqrt{2}}\begin{pmatrix}
    \frac{\Phi(\vec{k})}{|\Phi(\vec{k})|} & -\frac{\Phi(\vec{k})}{|\Phi(\vec{k})|}\\
    1 & 1\\
    \end{pmatrix}.
\end{equation}
In its general form, the hybridization term $\Hig$ couples the impurity to the graphene momentum state $\kk$ in subband $\alpha$ as
\begin{equation}
    \Hig = V\sum_{s,\alpha}\sum_{\vec{k}}\left\{\Theta_{I}^\alpha(\vec{k})\,d^\dagger_{I,s} c_{\alpha,s}(\kk) + \text{H.\ c.} \right\},
\end{equation}
with $V$ a real hopping constant. Encoded in $\Theta_{I}^\alpha(\kk)$ are the symmetry properties of the impurity orbital. In this article we study four of the most common atomic impurities found in real graphene samples: top--site ($\mathrm{TOP}$) and hollow--site ($\mathrm{HS}$) adatoms, and symmetric ($\mathrm{VAC}$) and reconstructed ($\mathrm{REC}$) single vacancies. Their corresponding couplings $\Theta_I^\alpha(\kk)$ are evaluated in Appendix \ref{app:impurities} using the real--space form of the graphene Hamiltonian, and their symmetry properties are discussed below.

$\mathrm{TOP}$ adatoms are the simplest of these impurity types. They sit on top of and couple exclusively to a single carbon atom [Fig.\ \ref{fig:impurities}(a)], thus singling out one of the sublattices and locally breaking inversion symmetry. Their point--like isotropic nature makes $\mathrm{TOP}$ adatoms couple equally to all graphene momenta as
\begin{equation}\label{eq:vTOP}
    \ThetaT^\pm(\vec{k}) = \frac{1}{\sqrt{2}}.
\end{equation}
Examples of $\mathrm{TOP}$ impurities have been recently reported in experiments with hydrogen atoms chemisorbed onto graphene.\cite{lin_nanolett_2015,Gonzalez-Herrero2016}  Although the impurity states were found to extend over several lattice constants from the adsorption site, local breaking of inversion symmetry was observed, and the model (\ref{eq:vTOP}) may be used as a first approximation.

\begin{figure}[t]
\begin{center}
\includegraphics[width=\columnwidth]{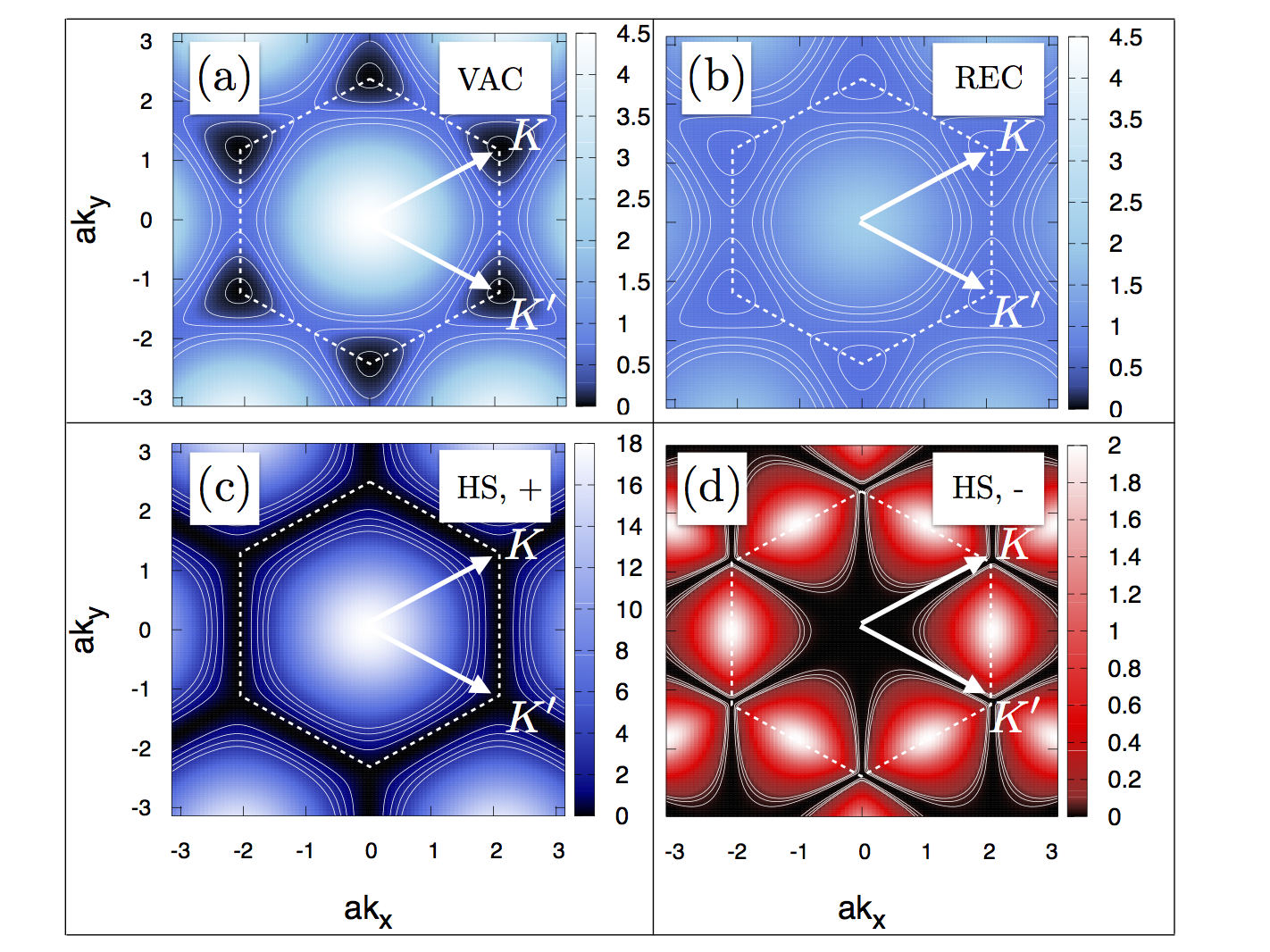}
\caption{(Color online) Scattering rate $|\Theta_I^{\alpha}(\kk)|^2$ for (a) symmetric and (b) reconstructed vacancies (both subbands), and for hollow--site adsorbates [(c) upper ($\alpha = +$) and (d) lower ($\alpha = -$) subband]. White equipotential contours are added to show topographic details, and the Brillouin zone is indicated as a dashed hexagon. Zeros appear at the $K$ and $K'$ points for $\mathrm{VAC}$ and $\mathrm{HS}$ impurities due to $C_{3v}$ symmetry, whereas inversion--symmetry--protected nodes appear only for $\mathrm{HS}$ impurities [panels (c) and (d)]. $C_{3v}$ symmetry is present in average for an ensemble of $\mathrm{REC}$ impurities, but the scattering rate is finite for all momenta.}
\label{fig:thetasq}
\end{center}
\end{figure}

On the other hand, more usual magnetic impurity candidates, such as transition metals, tend to adsorb in the hollow site.\cite{eelbo_prl_2013,donati_prl_2013,donati_prl_2014} $\mathrm{HS}$ adatoms with $s$ or $d_{z^2}$ valence orbitals will couple equally to both sublattices, and thus preserve both the inversion and point symmetries [Fig.\ \ref{fig:impurities}(b)]. The resulting coupling functions
\begin{equation}\label{eq:vHS}
    \ThetaH^\pm(\vec{k}) = \frac{1}{\sqrt{2}}\left[ \Phi^*(\vec{k}) \pm \frac{\Phi^2(\vec{k})}{|\Phi(\vec{k})|} \right],
\end{equation}
vanish at the $K$ and $K'$ points, and possess $C_{3v}$ symmetry about those point inherited from the function $\Phi(\kk)$, as shown in Figs.\ \ref{fig:thetasq}(c)  and \ref{fig:thetasq}(d). Moreover, inversion symmetry guarantees the presence of ``nodes'' in the coupling function---branches of graphene momenta that remain decoupled from the impurity degrees of freedom.\cite{ruiz-tijerina_2016} As we will discuss in Sec.\ \ref{sec:resistivity}, the zeros of the coupling function play a determinant role in the impurity contribution to the system transport properties.

Vacancies introduce localized midgap states in graphene\cite{Pereira06,Pereira08,yuan_prb_2010,lehtinen_prb_2010} that can develop magnetic moments\cite{Cazalilla12,Miranda14,Miranda2016} as a result of Coulomb charging--energy effects.\cite{PereiraPRL2007} In fact, recent studies show that such charging energies can be large\cite{Miranda2016} (of order $0.5\,\mathrm{eV}$), leading to the formation of effective $\pi$-like magnetic moments. In the case of symmetric vacancies, experiments have demonstrated charge accumulation at the vacancy site,\cite{andrei_natphys_2016} constituting the type of impurity we label as $\mathrm{VAC}$. To a first approximation, this impurity will couple identically and exclusively to its three nearest neighbors, belonging to the opposite sublattice\cite{ducastelle_prb_2013} [Fig.\ \ref{fig:impurities}(c)]. As a result, a $\mathrm{VAC}$ impurity will possess $C_{3v}$ symmetry but break local inversion symmetry. This is encoded in the coupling
\begin{equation}\label{eq:vVAC}
    \ThetaV^\pm(\vec{k}) = \frac{\Phi(\vec{k})}{\sqrt{2}},
\end{equation}
which is $C_{3v}$-symmetric about, and vanishes at the $K$ and $K'$ points [Fig.\ \ref{fig:thetasq}(a)], but lacks the inversion--symmetry nodes displayed by $\mathrm{HS}$ impurities.

Finally, vacancies with bond reconstruction\cite{barbary_prb_2003,ma_newjphys_2004,skowron_chemsocrev_2015} are the least symmetric of all cases, breaking both $C_{3v}$ and inversion symmetries. Bond reconstruction consists of a local deformation due to Jahn-Teller effects,\cite{barbary_prb_2003, kanao_jpsj_2012,palacios_prb_2012,lee_prb_2014,padmanabhan_prb_2016,valencia_2016} which allows a coupling between one carbon's $sp^2$ orbital and the $\pi$ orbitals belonging to the other two carbons surrounding the vacancy site---what we call the $\mathrm{REC}$ impurity [Fig.\ \ref{fig:impurities}(d)]. Placing the impurity orbital explicitly at $a\hat{\vec{u}}_j$ from the vacancy center, the coupling is given by
\begin{equation}\label{eq:vREC}
    \Theta_{\mathrm{REC},j}^\pm(\vec{k}) = \frac{\exp{ia\vec{k}\cdot\hat{\vec{u}}_j}}{\sqrt{2}}\sum_{l\ne j}\exp{-ia\vec{k}\cdot\hat{\vec{u}}_l}.
\end{equation}
Anticipating our transport discussion, one may conclude that such breaking of rotational symmetry should introduce dramatic anisotropy in the resistivity tensor $\varrho^{ij}$. In an ensemble, however, the $\mathrm{REC}$ impurities will occur at $j=1,\,2$ and $3$ with equal probability. In the dilute limit, where second and higher--order coherent scattering processes are neglected, the $\mathrm{REC}$ impurity distribution can be represented by the average scattering rate
\begin{equation}\label{eq:recav}
\begin{split}
    |\ThetaR(\vec{k})|^2 &= \frac{1}{3}\sum_{j=1}^3|\Theta_{\mathrm{REC},j}(\vec{k})|^2\\ &= 1+|\Phi(\vec{k})|^2 - \frac{2}{3}\real{\Phi^2(\vec{k})}.
\end{split}
\end{equation}
Eq.\ (\ref{eq:recav}) shows that the spatial averaging recovers $C_{3v}$ symmetry, but destroys the quantum interference leading to the zeros at $K$ and $K'$ [Fig.\ \ref{fig:thetasq}(b)].

At low energies the graphene sample properties depend only on the momentum states $\kp$ close to the Dirac points ($\kappa \ll K$). The low--energy theory is obtained by making
\begin{equation}
    \psi_s(\kp) = \begin{pmatrix} a_s(\vec{K}+\kp)\\b_s(\vec{K}+\kp)\\b_s(\vec{K}'+\kp)\\a_s(\vec{K}'+\kp) \end{pmatrix},\,c_s(\kp) = \begin{pmatrix} c_{+,s}(\vec{K}+\kp)\\c_{-,s}(\vec{K}+\kp)\\c_{-,s}(\vec{K}'+\kp)\\c_{+,s}(\vec{K}'+\kp)\end{pmatrix},
\end{equation}
and $h(\kk) \rightarrow \hbar v_F\,\tau^3 \vsig \cdot \kp$, where $\hbar v_F = 3a t/2$, $\tau^3$ is the third Pauli matrix in valley space, and $\vsig$ is the vector of Pauli matrices in sublattice space. Naturally, this model produces four bands $\epsilon_1(\kappa) = \epsilon_4(\kappa) = -\epsilon_2(\kappa) = -\epsilon_3(\kappa) = \hbar v_F \kappa$, corresponding to the valence and conduction Dirac cones at the $K$ and $K'$ points.

In this approximation the influence of graphene electrons with energy $\omega$ on the impurity is determined by the hybridization function
\begin{equation}\label{eq:hyb}
    \Gamma_I(\omega) = \pi V^2\sum_{\mu=1}^4\sum_{\kappa < \kappa_c}|\Theta_I^\mu(\kp)|^2\delta(\omega - \epsilon_{\mu}(\kappa)),
\end{equation}
where $\kappa_c = 3^{-3/4}(8\pi)^{1/2}a^{-1}$ is the Debye momentum cutoff,\cite{castro-neto_rmp_2009} and $\Theta_I^\mu(\kp)$ are the elements of the $4 \times 1$ coupling matrix
\begin{equation}
    \Theta_I(\kp) = \begin{pmatrix} \Theta_I^+(\vec{K}+\kp)\\ \Theta_I^-(\vec{K}+\kp)\\ \Theta_I^-(\vec{K}'+\kp)\\ \Theta_I^+(\vec{K}'+\kp) \end{pmatrix}.
\end{equation}
The hybridization functions corresponding to the couplings (\ref{eq:vTOP}) through (\ref{eq:vREC}) are 
\begin{subequations}\label{eq:hyb}
\begin{equation}\label{eq:hybT}
    \GammaT(\omega) = \Gamma_0\abs{\frac{\omega}{D}},
\end{equation}
\begin{equation}\label{eq:hybH}
    \GammaH(\omega) = \GammaV(\omega) = 4\pi\sqrt{3} \Gamma_0\abs{\frac{\omega}{D}}^3,
\end{equation}
\begin{equation}\label{eq:hybR}
    \GammaR(\omega) = \Gamma_0\left[ \abs{\frac{\omega}{D}} + 6\pi\sqrt{3}\abs{\frac{\omega}{D}}^3  \right],
\end{equation}
\end{subequations}
where $\Gamma_0 = 2\pi V^2/D$, and $D = \hbar v_F \kappa_c$ is the half--bandwidth of the graphene dispersion. Notice that the four impurity types can be grouped into two categories, depending on the low--energy behavior of their hybridization functions: the non-symmetric $\mathrm{TOP}$ and $\mathrm{REC}$ impurities, which couple to low--energy graphene states as $|\omega|$, and the highly--symmetric $\mathrm{HS}$ and $\mathrm{VAC}$ impurities, which do so as $|\omega|^3$, a result previously obtained in Ref.\ [\onlinecite{uchoa_prl_2011}]. For simplicity, in the following sections we will explicitly discuss $\mathrm{TOP}$ and $\mathrm{HS}$ impurities as representatives of their corresponding categories, with the express understanding that $\mathrm{REC}$ and $\mathrm{VAC}$ impurities, respectively, display qualitatively similar behaviors.

\section{Numerical results}\label{sec:results}
For $U \ne 0$ the Hamiltonian $H$ describes a system with strong spin correlations between the impurity and the graphene band. In the specific case of charge neutrality ($\mu \!=\! 0$), it corresponds to the pseudogap Anderson model,\cite{gonzalez-buxton_prb_1998} where the effective density of states coupled to the impurity level vanishes at the Fermi level as a power law $|\omega/D|^r$. Eqs.\ (\ref{eq:hyb}) give $r\!=\!1$ for $\mathrm{TOP}$ and $\mathrm{REC}$ impurities and $r\!=\!3$ for $\mathrm{HS}$ and $\mathrm{VAC}$ impurities.

In general, this problem cannot be solved analytically in closed form. Instead, we used Wilson's numerical renormalization group (NRG),\cite{wilson_1975,krishna_murthy_1980_1,krishna_murthy_1980_2,bulla_rmp_2008} adapted for a generic density of states following Ref.\ [\onlinecite{gonzalez-buxton_prb_1998}]. NRG is generally regarded as the method of choice for studying strongly correlated quantum impurity problems. It consists of numerically diagonalizing the Hamiltonian $H$ by logarithmically discretizing the energy--dependent hybridization functions [Eqs.\ (\ref{eq:hyb})] into energy bins $D\Lambda^{-(n+1)} < \pm \omega_{\pm,n} < D\Lambda^{-n}$, with $\Lambda > 1$ the discretization parameter and $n \ge 0$ an integer. This discretization scheme prevents artificially introducing an energy scale into the problem that may obscure any emergent scales, such as the Kondo temperature. The states belonging to bins $(\pm,n)$ are mapped onto a chain of fermionic states with local energies and hopping terms of order $D\Lambda^{-n/2}$ that fall exponentially with $n$. This so--called Wilson chain is coupled to the impurity site ($n \!=\! -1$) and diagonalized iteratively, deriving at every iteration $n$ an effective free energy valid for temperatures of order $D\Lambda^{-n/2}$.

In the following sections we present NRG results for the spectral density and thermodynamic properties of the different impurity types. The temperature--dependent charge and magnetic susceptibility will be used to accurately characterize the system ground state, unveiling a quantum phase transition (QPT) as a function of the impurity energy $\varepsilon_I$ in the absence of a back gate ($\mu = 0$), and the screening of the impurity spin through Kondo correlations for $\mu\!\ne\!0$. The impurity spectral density will be used to determine the low--temperature electronic transport properties of the graphene sample based on the formalism presented in Ref.\ [\onlinecite{ruiz-tijerina_2016}]. All calculations shown below were carried out with discretization factor $\Lambda = 2.5$, retaining approximately $1024$ states after each iteration. The spectral densities presented in Sec.\ \ref{sec:SpecDens} were evaluated using the density--matrix NRG method (DM-NRG).\cite{hofstetter_prl_2000}

\subsection{Quantum phase transition}\label{sec:QPT}
\begin{figure}[t]
\begin{center}
\includegraphics[width=\columnwidth]{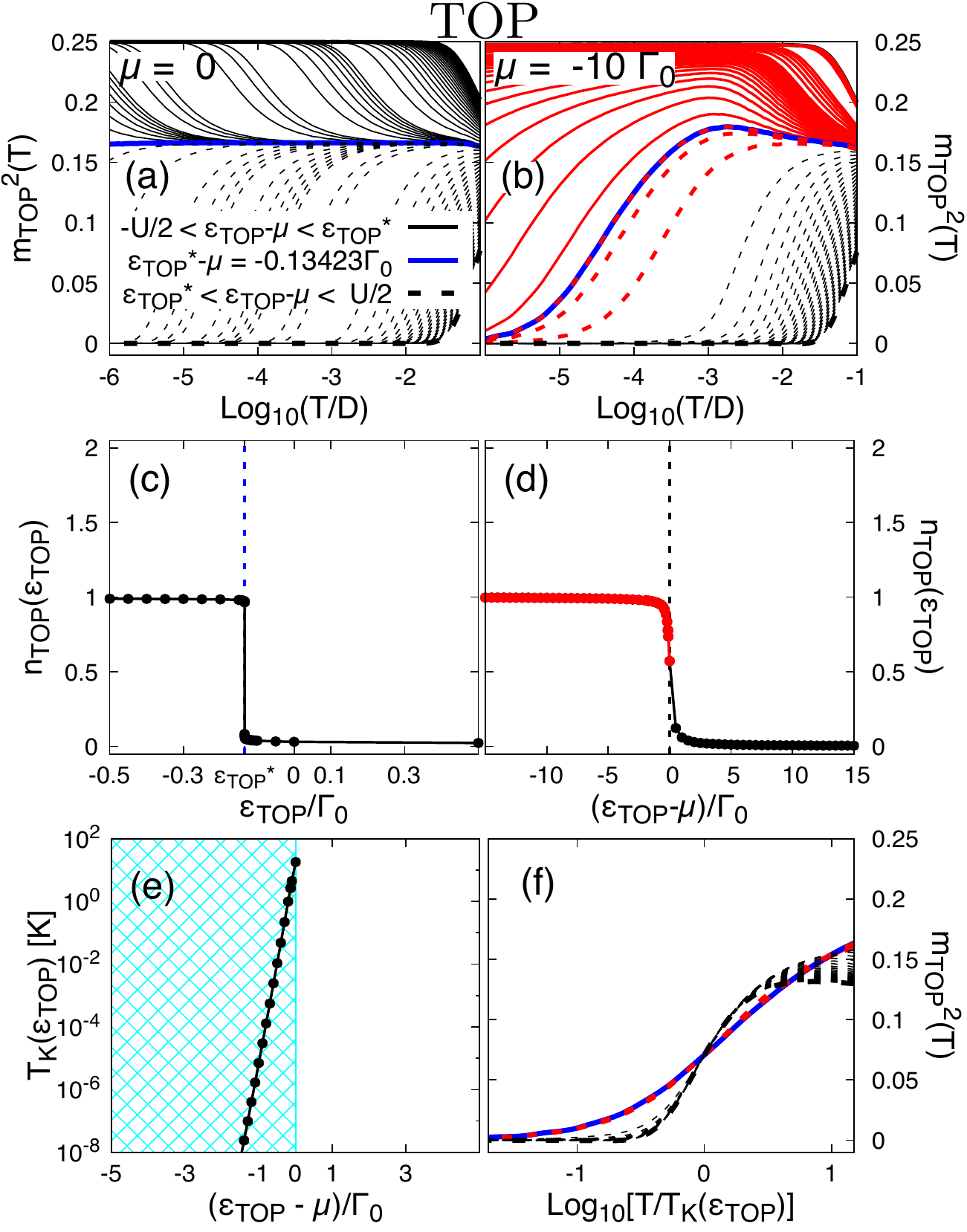}
\caption{(Color online) TOP--impurity magnetic moment squared as a function of temperature [(a) and (b)] and ground--state impurity charge as a function of $\eeT$ [(c) and (d)], for $\mu = 0$ and $\mu = -10\Gamma_0$, respectively. For $\mu = 0$, a phase transition from LM ($m^2_{\mathrm{TOP}} = 1/4$, $n_{\mathrm{TOP}} = 1$) to EO ($m_{\mathrm{TOP}}^2 = 0$, $n_{\mathrm{TOP}}=0$) occurs at critical energy $\eeT^* \approx -0.134225\,\Gamma_0$. (e) Away from charge neutrality ($\mu = -10\,\Gamma_0$) the impurity moment is Kondo-screened for $\eeT-\mu < 0$ [red curves and points in (b) and (d)], with Kondo temperatures $T_K \lesssim 10\, \mathrm{K}$. (f) Kondo universality (dashed red and solid blue curves) is observed for $\eeT- \mu \lesssim 0$ when the temperature is rescaled as $T/T_K(\eeT)$. A crossover to mixed valence occurs when the impurity level goes above the Fermi energy, indicated by a departure from universality. Parameters: $\Gamma_0 = U/30 = 0.01 D \sim 100\,\mathrm{meV}$.}
\label{fig:thermo_TOP}
\end{center}
\end{figure}
\begin{figure}[t]
\begin{center}
\includegraphics[width=\columnwidth]{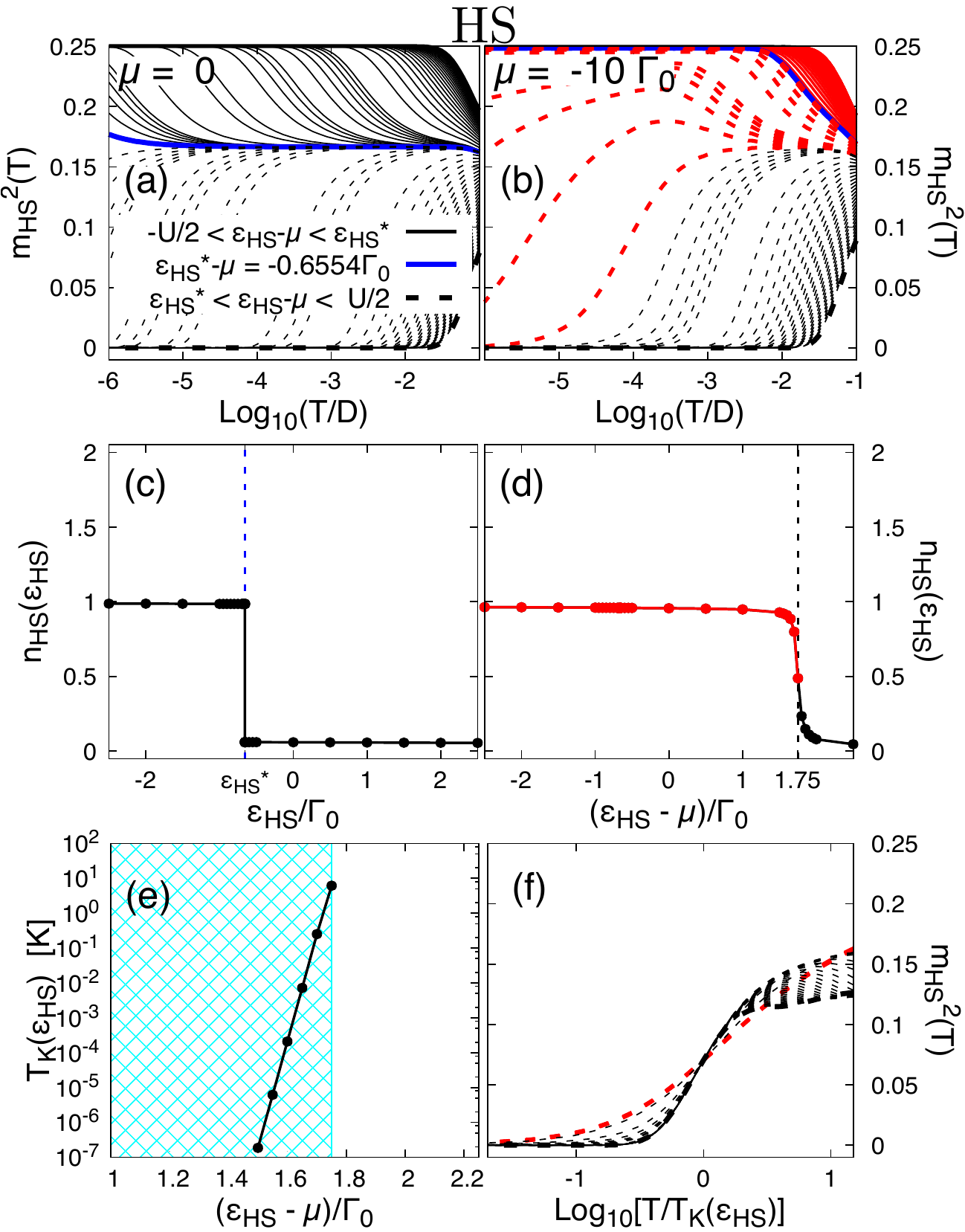}
\caption{(Color online) HS--impurity magnetic moment squared as a function of temperature [(a) and (b)], and ground--state impurity charge as a function of $\eeH$ [(c) and (d)], for  $\mu = 0$ and $\mu = -10 \Gamma_0$, respectively. A transition from LM to EO takes place for $\eeH > \eeH^* \approx -0.6554 \Gamma_0$ in the case of $\mu = 0$, whereas for $\mu = -10 \Gamma_0$ there is a Kondo ground state [red curves and points in (b) and (d)] for $\eeH-\mu \lesssim 1.7\Gamma_0$, with Kondo temperatures $T_K \lesssim 10 \,\mathrm{K}$ [panel (e)]. The sudden impurity charge drop in (d) at $\eeH \approx 1.75 \Gamma_0$ indicates a strong renormalization of the impurity level toward negative energies for finite $\mu$. Parameters: $\Gamma_0 = U/30 = 0.01 D \sim 100\,\mathrm{meV}$.}
\label{fig:thermo_HS}
\end{center}
\end{figure}

Fig.\ \ref{fig:thermo_TOP} compares the behavior of a $\mathrm{TOP}$ impurity's magnetic moment squared\footnote{In dimensionless form, $m_I^2(T) = T\chi_I(T)/(g\mu_B)^2$, with $\chi_I(T)$ the contribution of impurity $I$ to the magnetic susceptibility, $\mu_B$ the Bohr magneton, and $g$ the $g$ factor.} $m_{\mathrm{TOP}}^2(T)$ and ground--state impurity level occupation $n_{\mathrm{TOP}} \equiv \langle d_{\mathrm{TOP},s}^\dagger d_{\mathrm{TOP},s} \rangle$ for zero and non--zero values of the chemical potential. For $\mu\!=\!0$ a realization of the $r=1$ pseudogap Anderson model is obtained,  known\cite{fritz_vojta_prb_2004,fritz_vojta_prb_2_2004,fritz_vojta_kondo_in_graphene} to display critical behavior for a given impurity energy $\eeT^*<0$. This marks a QPT between a local--moment (LM) ($\eeT < \eeT^*$) and an empty--orbital (EO) phase ($\eeT > \eeT^*$). The LM phase consists of a ground state where the impurity is charged with a single electron and behaves as a free spin $1/2$, characterized by $m_{\mathrm{TOP}}^2(0)\!=\!1/4$ [solid lines in Fig.\ \ref{fig:thermo_TOP}(a)] and $n_{\mathrm{TOP}}(0)\!=\!1$  [Fig.\ \ref{fig:thermo_TOP}(c), left]. In contrast, in the EO phase the impurity is depleted below some transition temperature, leading to $m_{\mathrm{TOP}}^2(0)\!=\!0$ [dashed lines in Fig.\ \ref{fig:thermo_TOP}(a)] and $n_{\mathrm{TOP}}(0)\!=\!0$ [Fig.\ \ref{fig:thermo_TOP}(c), right].

The situation is markedly different for a finite chemical potential. Figs.\ \ref{fig:thermo_TOP}(b) and (d) show $m_{\mathrm{TOP}}^2(T)$ and $n_{\mathrm{TOP}}(T)$, respectively, for $\mu\!=\!-10 \Gamma_0$. No QPT is observed in this case; instead, there is a smooth crossover from an EO phase ($\eeT>\mu$) to a Kondo regime ($\eeT<\mu$), the expected behavior for the usual (metallic) Anderson model. For each $\eeT$ we extracted the Kondo temperature $T_K(\eeT)$ from the corresponding $m_{\mathrm{TOP}}^2(T)$ vs.\ $T$ curve as the temperature for which the crossover from LM to Kondo singlet is completed. Following Wilson's convention,\cite{wilson_1975} we estimate $m_{I}^2(T_K)=0.0707$. Remarkably, Kondo temperatures as high as $10\,\mathrm{K}$ can be obtained for $\mathrm{TOP}$ impurities with these parameters, as shown in Fig.\ \ref{fig:thermo_TOP}(e). Rescaling temperature as $T/T_K(\eeT)$ collapses all the $m_{\mathrm{TOP}}^2(T)$ curves in the Kondo regime into a single universality curve, shown with red dashed lines in Fig.\ \ref{fig:thermo_TOP}(f). Notice that the curve for $\eeT - \mu = \eeT^*$ [blue curve in \ref{fig:thermo_TOP}(b) and \ref{fig:thermo_TOP}(f)] falls in the Kondo regime when $\mu \ne 0$. When $\eeT > \mu$ the impurity is discharged and the Kondo effect is no longer possible, leading to an EO ground state instead. The crossover temperature cannot be interpreted as a Kondo temperature in this case, and after rescaling the $m_{\mathrm{TOP}}^2(T)$ curves no longer follow Kondo universality [black dashed curves in Fig.\ \ref{fig:thermo_TOP}(f)].

\begin{figure*}[t]
\begin{center}
\includegraphics[width=1.95\columnwidth]{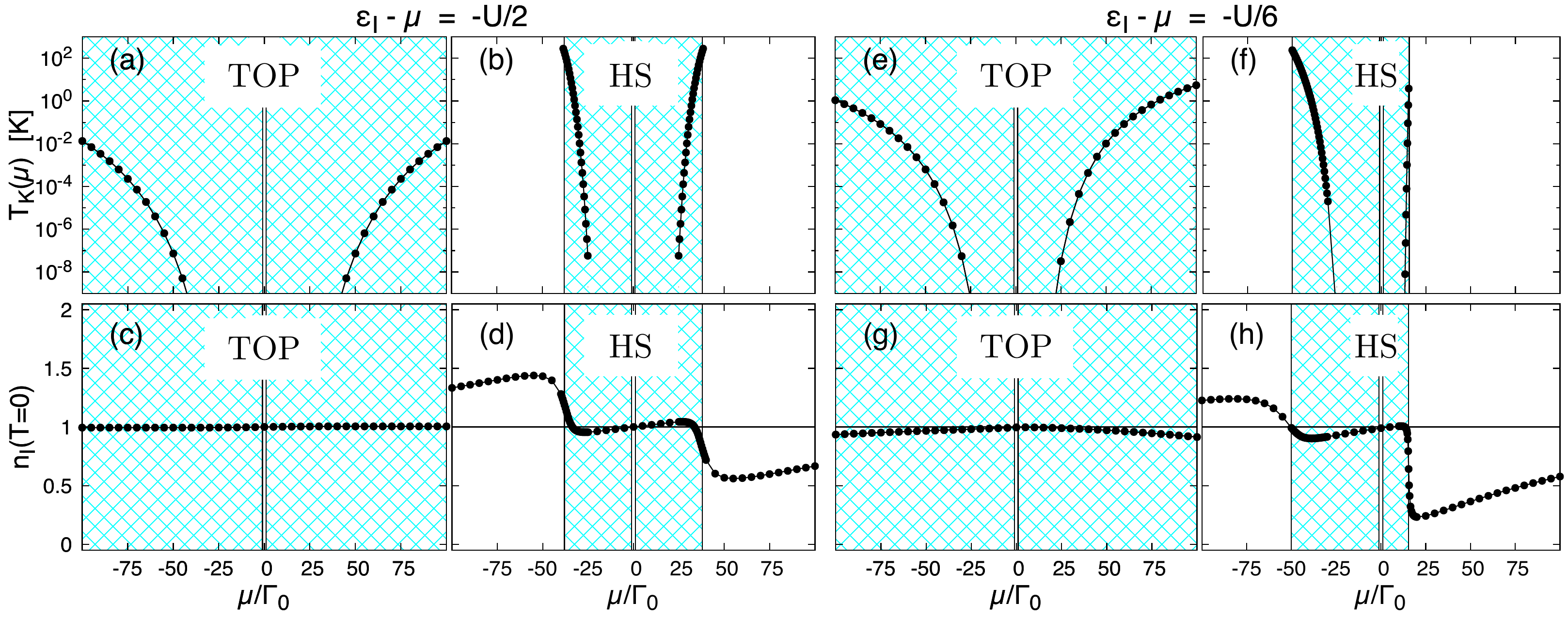}
\caption{Kondo temperature and ground--state impurity charge for $\mathrm{TOP}$ and $\mathrm{HS}$ impurities as functions of the chemical potential $\mu$, for $\ee_I-\mu = -U/2$ [panels (a) through (d)] and $\ee_I-\mu = -U/6$ [panels (e) through (h)]. In both cases $\mathrm{TOP}$ impurities undergo Kondo screening for all finite $\mu$; the LM phase remains down to zero temperature for $\mu=0$, where $T_K = 0$ [(a) and (e)]. (c) For $\mathrm{TOP}$ adatoms the impurity charge remains at $n_{\mathrm{TOP}} = 1$ quite independently of $\mu$, even for an asymmetric [$2(\eeT-\mu)+U\ne0$] impurity (g). This is not the case for $H$ impurities, whose charge strongly deviates from $n_{\mathrm{HS}} = 1$ antisymmetrically (asymmetrically) with $\mu$ for $2(\eeH-\mu)+U=0$ [$2(\eeH - \mu) +U\ne0$] [(d) and (h)]. Kondo spin correlations can only exist when the $\mathrm{HS}$ adatom has a well--defined charge of $n_{\mathrm{HS}} \approx 1$, so the Kondo effect is limited to the shaded regions of panels (b), (d), (f) and (h). Parameters: $\Gamma_0 = U/30 = 0.01 D \sim 100\,\mathrm{meV}$.}
\label{fig:TK_and_charge}
\end{center}
\end{figure*}

A qualitatively similar picture is found for ${\mathrm{HS}}$ impurities, as seen in Fig.\ \ref{fig:thermo_HS}. Setting $\mu\!=\!0$ we have a realization of the $r\!=\!3$ pseudogap Anderson model, which displays a LM-EO QPT with $\eeH^* \approx -0.6554\Gamma_0$ [Figs.\ \ref{fig:thermo_HS}(a) and (c)], a critical energy about 5 times larger than its $r\!=\!1$ counterpart. For $\mu\!=\!-10\Gamma_0$, Fig.\ \ref{fig:thermo_HS}(b) shows a crossover from the EO phase to the Kondo regime as $\eeH - \mu$ is tuned toward zero from the positive side. Surprisingly, the Kondo regime begins at unusually large positive values of the bare impurity level, $\eeH - \mu\!\approx\!1.7\,\Gamma_0$, as shown in Fig.\ \ref{fig:thermo_HS}(d), with Kondo temperatures reaching values of order $10\,\mathrm{K}$ [Fig.\ \ref{fig:thermo_HS}(e)]. The persistence of Kondo correlations was confirmed by scaling $T/T_K(\eeH)$ and verifying that the $m_{\mathrm{HS}}^2(T/T_K(\eeH))$ curves collapsed into the typical Kondo universality curve up to $\eeT - \mu \approx 1.7\,\Gamma_0$. This is shown by the red dashed lines in Fig.\ \ref{fig:thermo_HS}(f). The Kondo temperature is lowered as the impurity level is shifted toward more negative values [Fig.\ \ref{fig:thermo_HS}(b)]. By the time $\eeH - \mu = \eeH^*$ [blue curve in Fig.\ \ref{fig:thermo_HS}(b)] the Kondo temperature is so low that, for all practical purposes, the system ground state must be considered LM.

We believe these large value of $\eeH^*$ and the unexpected persistence of Kondo physics up to $\eeH - \mu\!\approx\!1.7\,\Gamma_0$ are the result of a two--stage process, wherein the impurity level and hybridization are strongly renormalized by interactions at the higher energy scales, and the magnetic moment then undergoes Kondo screening at lower energies. The QPT for $\mu\!=\!0$ and $r\!=\!1$ is associated with a quasiparticle level crossing,\cite{fritz_vojta_prb_2004} where the bare impurity level is shifted to positive values, and the shifted level crosses zero when $\eeT\!=\!\eeT^*<0$. The situation is similar for $r>1$, with logarithmic corrections to the level crossing\cite{gonzalez-buxton_prb_1998,fritz_vojta_prb_2_2004} that may justify  that $|\eeH^*| > |\eeT^*|$. For finite $\mu$, on the other hand, a renormalized impurity level $\tilde{\ee}_{\mathrm{HS}} < \eeH$ before the onset of Kondo correlations would justify the persistence of Kondo physics up to $\eeH-\mu > 0$ for $H$ impurities.

The high Kondo temperatures calculated for both $\mathrm{TOP}$ and $\mathrm{HS}$ impurities ($T_K \sim 10\,\mathrm{K}$) are encouraging from an experimental point of view. Previous studies, such as Refs.\ [\onlinecite{uchoa_prl_2011}] and [\onlinecite{po-wei-lo_prb_2014}], focused on $\mathrm{TOP}$ and $\mathrm{REC}$ impurities, arguing that Kondo temperatures for $\mathrm{HS}$ or $\mathrm{VAC}$ impurities would be much lower, and thus experimentally inaccessible. This is a reasonable expectation, given the well--known fact that, in both metallic\cite{haldane_1978} and gapped\cite{zitko_prb_2015} systems, the Kondo temperature depends exponentially on the hybridization strength near the Fermi level. However, our results show that Kondo temperatures comparable to those obtained for $\mathrm{TOP}$ and $\mathrm{REC}$ impurities are possible for sensible values of the carrier density [Figs.\ \ref{fig:TK_and_charge}(b) and \ref{fig:TK_and_charge}(f)].

The results presented in Fig.\ \ref{fig:TK_and_charge} demonstrate that Kondo temperatures of almost $100\,\mathrm{K}$ can be obtained with the application of a back gate.  For $\mathrm{TOP}$ or $\mathrm{REC}$ impurities, Figs.\ \ref{fig:TK_and_charge}(a) and \ref{fig:TK_and_charge}(e) show that the Kondo temperature can be increased by simply raising the chemical potential, and that higher values are reached for more asymmetric impurities [$2(\ee_{I}-\mu)+U \gg \Gamma_0$], consistent with the results of Fig.\ \ref{fig:thermo_TOP}(b). The situation is more subtle for $\mathrm{HS}$ or $\mathrm{VAC}$ impurities, where the Kondo regime exists only within specific ranges of $\mu$, as shown in Figs.\ \ref{fig:TK_and_charge}(b) and \ref{fig:TK_and_charge}(f), effectively setting a parameter--dependent upper limit for $T_K$.  This can be understood in terms of the impurity charge: A charge of $n_{\mathrm{HS}} \approx 1$ is required for the impurity to develop a magnetic moment of $|m_{\mathrm{HS}}| \approx 1/2$, which the graphene electrons may then screen through Kondo spin scattering. However, as shown in Figs.\ \ref{fig:TK_and_charge}(d) and \ref{fig:TK_and_charge}(h) for different $\mathrm{HS}$ impurity level energies, beyond certain values of the chemical potential the impurity charge significantly departs from that value, preventing the Kondo singlet from forming. This is in stark contrast to $\mathrm{TOP}$ or $\mathrm{REC}$ impurities, whose charges are quite independent of $\mu$ [Figs.\ \ref{fig:TK_and_charge}(c) and \ref{fig:TK_and_charge}(g)].

\subsection{Impurity spectral density}\label{sec:SpecDens}
\begin{figure*}[t]
\begin{center}
\includegraphics[width=1.98\columnwidth]{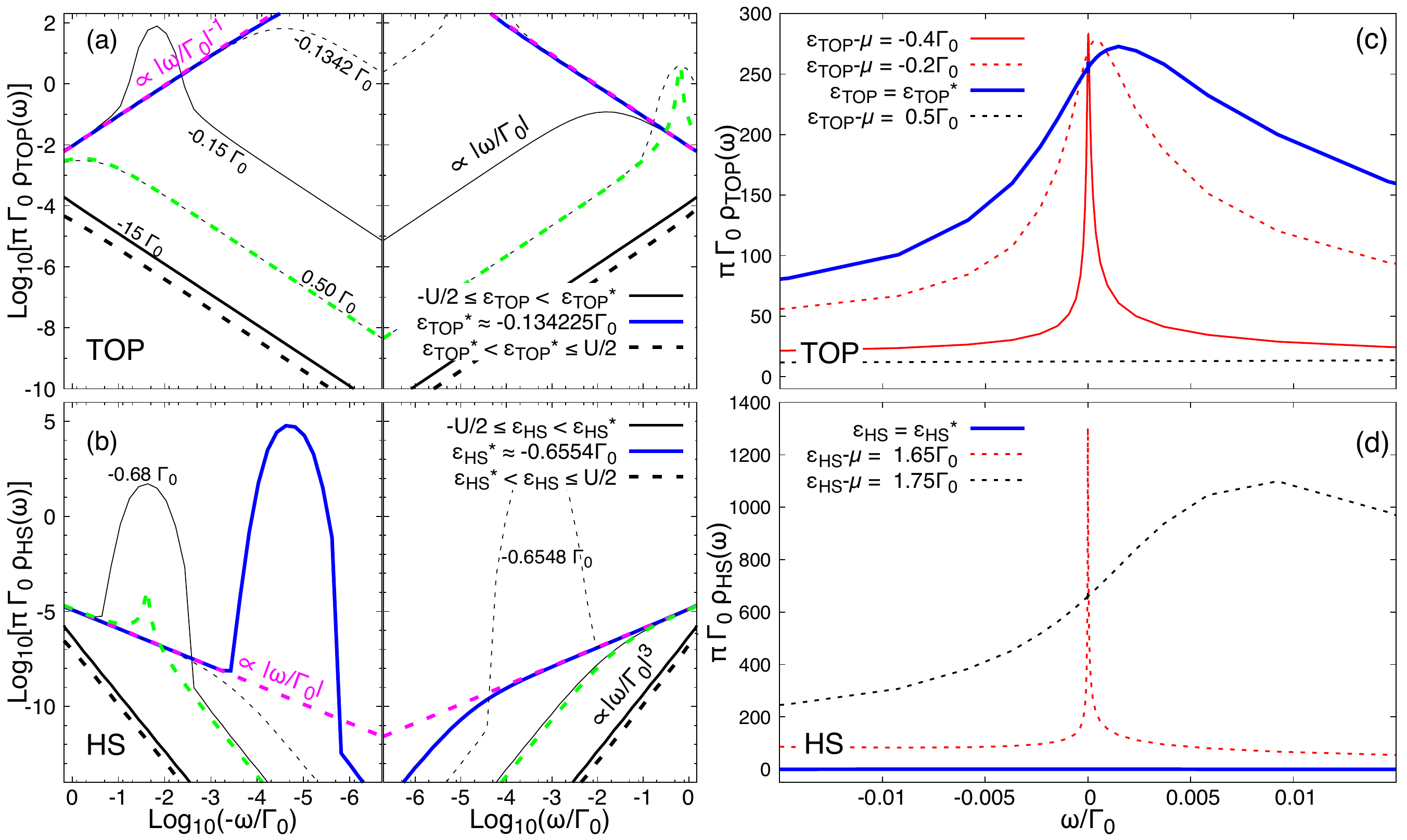}
\caption{(Color online) Spectral densities of $\mathrm{TOP}$ and $\mathrm{HS}$ impurities, for different impurity energies $\eeT$ and $\eeH$. (a) In charge--neutral graphene, the $\mathrm{TOP}$ spectral density displays $|\omega|^{-1}$ scaling at high energies, followed by a quasiparticle state at negative (positive) energy for $\eeT < \eeT^*$ ($\eeT > \eeT^*$), and linear scaling close to zero energy. At criticality ($\eeT = \eeT^*$), $|\omega|^{-1}$ scaling persists down to zero energy, and is fitted by Eq.\ (\ref{eq:crit_rho_T}) with $\tilde{\Gamma}_0 = 0.9\,\Gamma_0$ (pink dashed line). The same fitting is valid for $\eeT \ne \eeT^*$, as shown by the green dashed line. (b) $\mathrm{HS}$ impurities show a similar behavior, with linear scaling at high energies, a quasiparticle level at negative (positive) energy for $\eeH < \eeH^*$ ($\eeH > \eeH^*$), and $\rho_{\mathrm{HS}}(\omega) \sim D^2|\omega|^{-3}$ close to zero energy. The spectral density is fitted by Eq.\ (\ref{eq:crit_rho_H}) with $\tilde{\Gamma}_0 = 0.6\,\Gamma_0$ at and away from criticality (pink and green dashed lines, respectively). (c) For finite chemical potential, $\mathrm{TOP}$ impurities show a Kondo resonance (solid red line) for $\eeT < \mu$, which broadens as $\eeT \rightarrow \mu$ before the impurity charge drops [see Fig.\ \ref{fig:thermo_TOP}(d)] and the Kondo effect disappears. (d) For $\mathrm{HS}$ impurities, Kondo signatures are visible up to a positive (bare) impurity energy $\eeH-\mu \approx 1.7\, \Gamma_0$. Parameters: $\Gamma_0 = U/30 = 0.01 D \sim 100\,\mathrm{meV}$.}
\label{fig:spectral}
\end{center}
\end{figure*}

For charge--neutral graphene, the impurity spectral density $\rho_I(\omega) = -\pi^{-1}\imag{\rgd{\omega}}$, with $\rgd{\omega}$ the retarded impurity Green's function,\cite{zubarev} can be interpreted in terms of the renormalized--parameters picture described in Section \ref{sec:QPT}, and the corresponding result for the noninteracting case ($\mu\!=\!U\!=\!0$):\cite{ruiz-tijerina_2016} 
\begin{equation}\label{eq:spectral_analytical}
    \rho_I^0(\omega) = \frac{1}{\pi}\frac{\Gamma_I(\omega)}{[\omega - \ee_I]^2 + \Gamma_I^2(\omega)}.
\end{equation}
Setting $\ee_I\!=\!0$ and $|\omega| \ll \Gamma_0$, we obtain $\rho_{\mathrm{TOP}}^0(\omega) = D\Gamma_0\pi^{-1}(D^2+\Gamma_0^2)^{-1}|\omega|^{-1}$ and $\rho_{\mathrm{HS}}^0(\omega) \approx 4\sqrt{3}\Gamma_0D^{-3}|\omega|$. Similar power--law behaviors are obtained for the interacting case when $\eeT \approx \eeT^*$ and $\eeH \approx \eeH^*$, as shown with solid blue curves in Figs.\ \ref{fig:spectral}(a) and \ref{fig:spectral}(b), respectively. In both cases, when $\ee_I < \ee_I^*$ ($\ee_I > \ee_I^*$) a quasiparticle peak appears at a negative (positive) energy, and the spectral density vanishes exactly at the Fermi level with the same power law as the corresponding $\Gamma_I(\omega)$.  These results demonstrate that our estimated value for $\eeT^*$ is a good approximation down to energies $|\omega| \sim 10^{-5}\,\mathrm{meV}$, whereas our estimation for $\eeH^*$ is only good down to $|\omega| \sim 0.1\,\mathrm{meV}$, before the quasiparticle level becomes visible. In any case, the interacting problem (for $\mu\!=\!0$) can be understood to a good approximation in terms of the noninteracting picture, with the role of local interactions being simply to renormalize the impurity parameters as $\ee_I \rightarrow \tilde{\ee}_I = \ee_I - \ee_I^*$ and $\Gamma_0 \rightarrow \tilde{\Gamma}_0$. In criticality ($\tilde{\ee}_I = 0$) the spectral densities are given by
\begin{subequations}
\begin{equation}\label{eq:crit_rho_T}
    \rho_{\mathrm{TOP}}(\omega) = \frac{D\tilde{\Gamma}_0}{\pi(D^2 + \tilde{\Gamma}_0^2)} |\omega|^{-1},
\end{equation}
\begin{equation}\label{eq:crit_rho_H}
    \rho_{\mathrm{HS}}(\omega) = \frac{4\sqrt{3}\tilde{\Gamma}_0}{D^3} |\omega|.
\end{equation}
\end{subequations}
Remarkably, the NRG results of Fig.\ \ref{fig:spectral} can be nicely fitted by the non--interacting--case expressions (\ref{eq:crit_rho_T}) and (\ref{eq:crit_rho_H}), with $\tilde{\Gamma}_0$ the only free parameter, as shown in Fig.\ \ref{fig:spectral}. This reveals that interaction effects are essentially irrelevant (in the RG sense) at the critical point, and the physics of the transition can be described by non--interacting quasiparticles.

It is worthwhile mentioning that \textit{no Kondo peak appears for} $\mu\!=\!0$ on either side of the transition, regardless of the values of $\Gamma_0$ or $\ee_I$. By contrast, the impurity spectral densities for $\mu\!=\!-10\Gamma_0$ display the familiar Kondo peak\cite{abrikosov_1965,suhl_1965,nagaoka_1965} for $\eeT-\mu \lesssim 0$ in the case of $\mathrm{TOP}$ impurities, and for $\eeH-\mu \lesssim 1.7\,\Gamma_0$ for $\mathrm{HS}$ impurities, as shown by the red curves in Figs.\ \ref{fig:spectral}(c) and (d).

\subsection{Resistivity calculations}\label{sec:resistivity}
\begin{figure*}[t]
\begin{center}
\includegraphics[width=1.95\columnwidth]{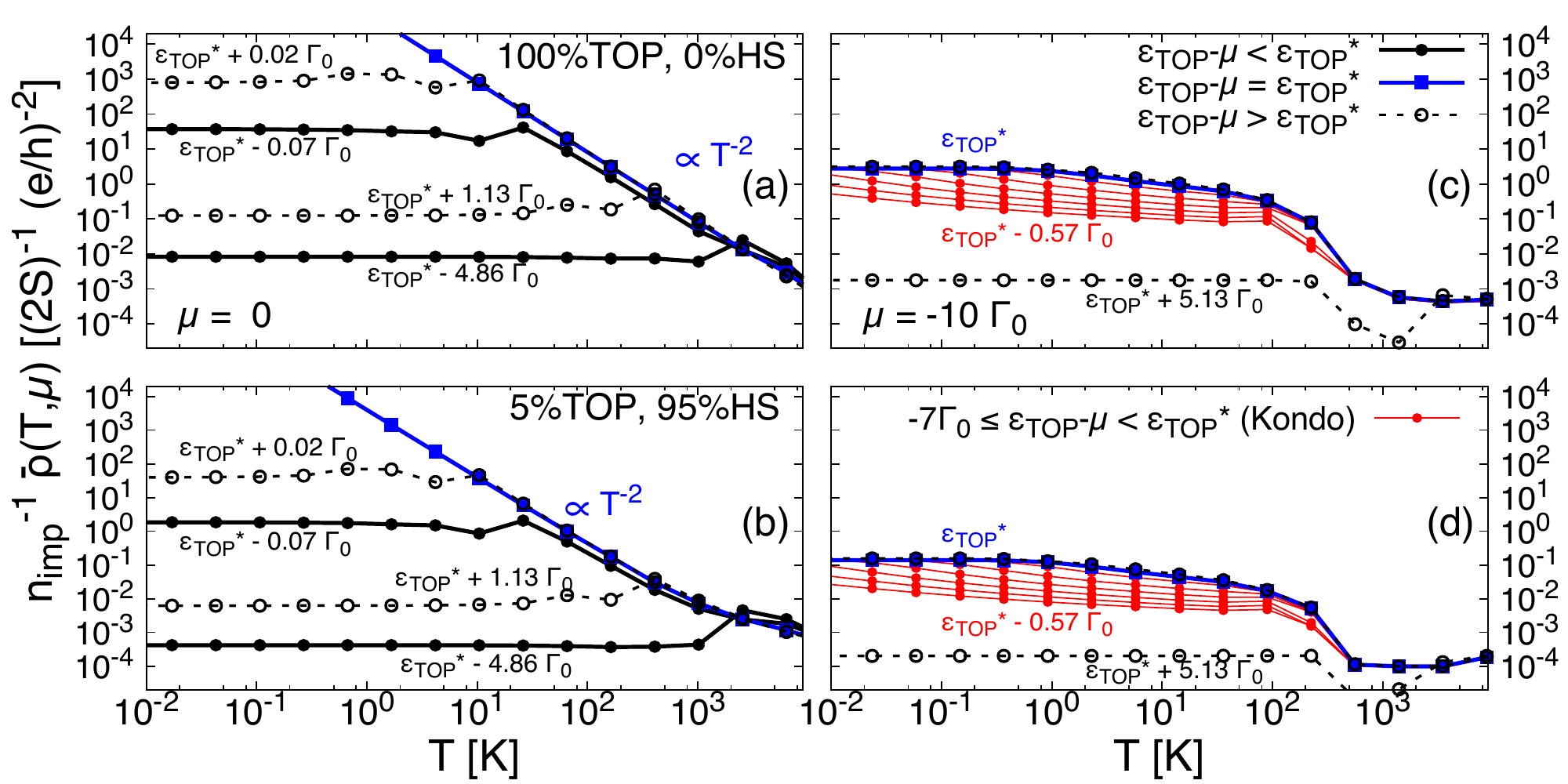}
\caption{(Color online) Resistivity as a function of temperature for samples with only $\mathrm{TOP}$ impurities $(n=0)$ [(a) and (c)] and a $5\%\,\mathrm{TOP}$/$95\%\,\mathrm{HS}$ $(n=0.95)$ mixture [(b) and (d)]. Individual values of $\eeT-\mu$ are indicated beside the corresponding curve. Panels (a) and (b) correspond to charge neutrality ($\mu\!=\!0$), with full (empty) circles representing impurity energies below (above) the critical point, $\varepsilon_I < \varepsilon_I^*$ ($\varepsilon_I > \varepsilon_I^*$). For both mixtures the critical impurity energy (blue squares) displays $T^{-2}$ temperature scaling, reminiscent of bound state formation by nonmagnetic resonant scatterers, whereas for $\eeT \ne \eeT^*$ the resistivity plateaus at a parameter--dependent value. Panels (c) and (d) show results for a finite chemical potential $\mu = -10\, \Gamma_0$. Red curves correspond to $\eeT$ values for which the Kondo effect appears at low temperatures. Notice that those curves plateau to the same resistivity value below their corresponding Kondo temperatures, $T_K(\eeT)$. Parameters: $\Gamma_0 = U/30 = 0.01 D \sim 100\,\mathrm{meV}$.}
\label{fig:res_vs_T_100T_00H}
\end{center}
\end{figure*}

The resistivity of a graphene sample with a dilute concentration of impurities\footnote{Our resistivity calculations neglect coherent multi--impurity scattering events, that are only relevant when the interimpurity distances are shorter than the graphene mean free path. This approximation is good for mesoscopic graphene samples, but may fall short for high-quality epitaxial graphene at low temperatures, where mean free paths of order 600 nm have been reported \cite{berger_science_2006}. Nonetheless, the qualitative resistivity features of highly-symmetric impurities discussed below, resulting from the vanishing impurity coupling to specific graphene states, do not depend on this approximation.} can be evaluated in terms of the single-impurity spectral density.\cite{mahan,costi_hewson_zlatic1994,ryu_prb_2007,cornaglia_prl_2009} We will consider real experimental situations, where a pure ensemble of a single impurity type is unlikely. In reality, evaporating adatoms onto the graphene sample will produce a mixture of $\mathrm{TOP}$ and $\mathrm{HS}$ impurities,\cite{eelbo_prl_2013} and creating vacancies through electron beam sputtering\cite{robertson_natcomm_2012} or ionic bombardment\cite{andrei_natphys_2016}  will produce a mixture of $\mathrm{REC}$ and $\mathrm{VAC}$ impurities. Therefore, we consider a low impurity density $n_{\text{imp}}$ consisting of a fraction $n$ of symmetric ($\mathrm{HS}$ or $\mathrm{VAC}$) impurities and a fraction $(1-n)$ of non--symmetric ($\mathrm{TOP}$ or $\mathrm{REC}$) impurities. 

Fig.\ \ref{fig:res_vs_T_100T_00H} shows the temperature dependence of the impurity contribution to the resistivity, for HS-TOP mixes (H-T) with different values of the local energy, assuming $\eeH\!=\!\eeT$. Two impurity fractions, $n=0$ (only $\mathrm{TOP}$ impurities) and $n = 0.05$ ($95\%$ of $\mathrm{HS}$ impurities), are considered for $\mu\!=\!0$ and $\mu\!=\!-10\Gamma_0$. Two transport regimes appear for $\mu\!=\!0$: when $\tilde{\ee}_{\mathrm{TOP}}\!\ne\!0$ ($\eeT \ne \eeT^*$), the transport is determined by normal impurity scattering, yielding a finite low--temperature resistivity plateau at $\sim (2\pi\tilde{\ee}_{\mathrm{TOP}}/\tilde{\Gamma}_0)^{-2}$. Then, for $\tilde{\ee}_{\mathrm{TOP}}=0$ ($\eeT = \eeT^*$) impurity criticality is signaled by full insulating behavior at zero temperature, with the scaling law $\varrho_{\mathrm{H-T}}(\eeT=\eeT^*,\mu=0) \sim (1-n)D^2\tilde{\Gamma}_0^2(D^2+\tilde{\Gamma}_0^2)^{-1}T^{-2}$, in complete correspondence with a nonmagnetic resonant scatterer (noninteracting impurity with $\ee_I=\mu=0$), where this behavior can be interpreted as the formation of an impurity bound state at the Fermi level.\cite{ferreira_prb_2011,asmar_prl_2014,ruiz-tijerina_2016}

As in the noninteracting case,\cite{ruiz-tijerina_2016} the impurity scattering in an H-T mixture comes exclusively from the $\mathrm{TOP}$ adatoms. Due to the symmetry--protected vanishing of $\ThetaH^{\pm}(\kk)$ at $K$, $K'$ and at the inversion--symmetry--protected nodes, $\mathrm{HS}$ impurities always allow for coherent transport through graphene momentum channels that remain decoupled from the impurity by symmetry, and thus do not contribute to the resistivity. This is the origin of the factor $(1-n)$ in the expression above.

\begin{figure}[t]
\begin{center}
\includegraphics[width=\columnwidth]{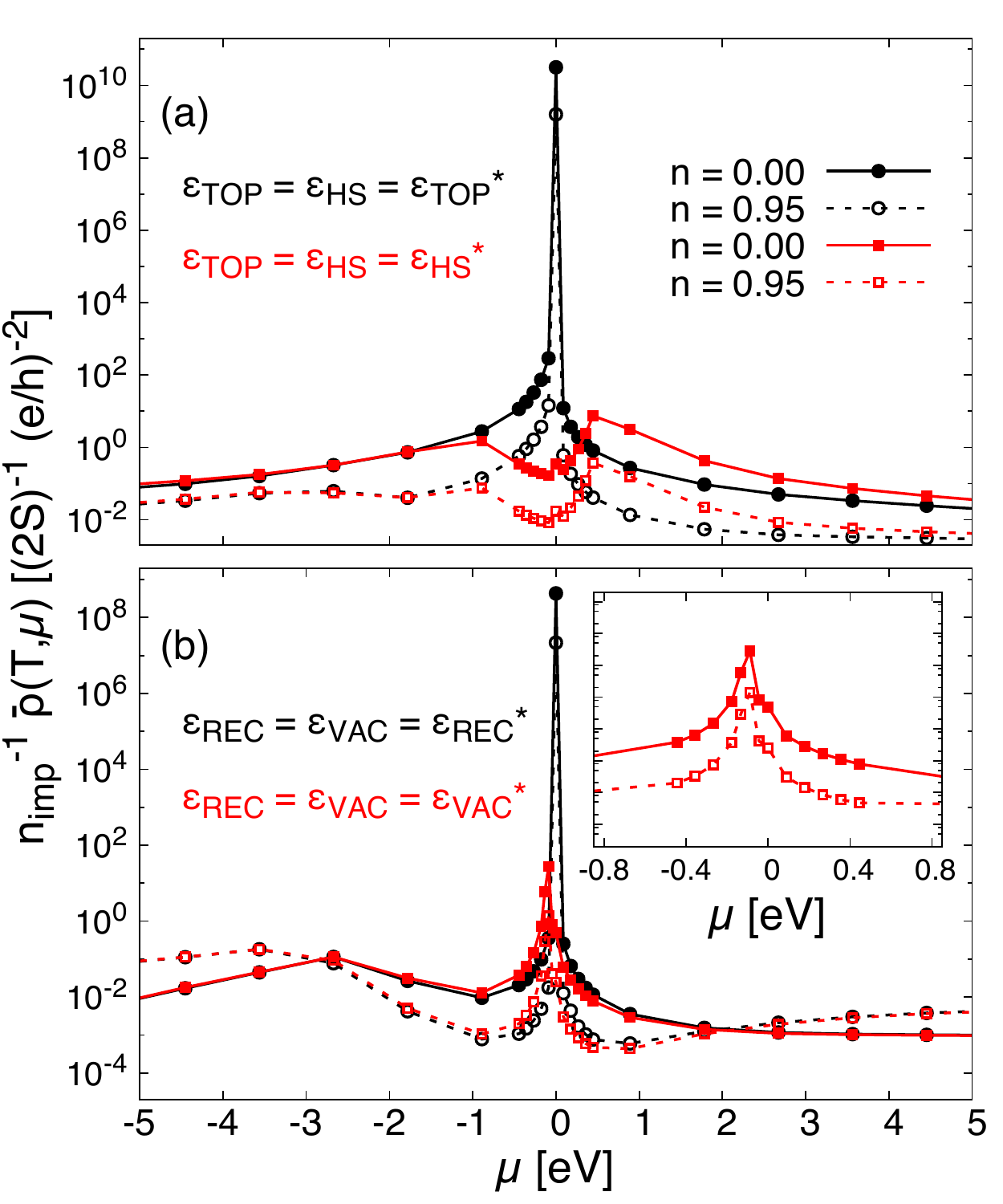}
\caption{(Color online) Low--temperature impurity contribution to the graphene sample resistivity for critical (a) top-- and hollow--site adatoms, and (b) reconstructed and symmetric vacancies. Solid curves and filled symbols correspond to samples with only $\mathrm{TOP}$ or $\mathrm{REC}$ impurities [(a) and (b), respectively], \emph{i.e.}, a symmetric impurity fraction $n\!=\!0$. Dashed curves and empty symbols correspond to $n\!=\!0.95$, or $95\%$ symmetric ($\mathrm{HS}$ or $\mathrm{VAC}$) and $5\%$ non--symmetric ($\mathrm{TOP}$ or $\mathrm{REC}$) impurities. In both cases, black curves and symbols represent an impurity energy $\eeT\!=\!\eeH =\!\eeT^*$, whereas red corresponds to $\eeT\!=\!\eeH =\!\eeH^*$. Parameters: $\Gamma_0 = U/30 = 0.01 D \sim 100\,\mathrm{meV}$, $T = 1.5\,\mathrm{mK}$.}
\label{fig:res_vs_mu}
\end{center}
\end{figure}

Fig.\ \ref{fig:res_vs_mu} shows low--temperature resistivity results for critical impurity energies as functions of the chemical potential, for two different symmetric--impurity ($\mathrm{HS}$ or $\mathrm{VAC}$) fractions $n = 0$ and $n=0.95$. When only critical $\mathrm{TOP}$ impurities ($\tilde{\ee}_{\mathrm{TOP}} = 0$) are present [full black circles in Fig.\ \ref{fig:res_vs_mu}(a)] the resistivity peaks sharply at zero chemical potential, in perfect analogy with the noninteracting case.\cite{ruiz-tijerina_2016} As the symmetric ($\mathrm{HS}$) impurity fraction is increased to $n=0.95$ for the same level energy [empty black circles in Fig.\ \ref{fig:res_vs_mu}(a)], a general $\sim 95\%$ drop is observed, while the resistivity profile is left unchanged, indicating that the only resistivity source are the $\mathrm{TOP}$ impurities in the ensemble. As a consequence, no criticality signature appears if the impurities are instead tuned to the $\mathrm{HS}$-impurity critical energy $\eeH^*$ [full red squares in Fig.\ \ref{fig:res_vs_mu}(a)]. The overall $\sim 95\%$ resistivity drop for $n=0.95$ [empty red squares in Fig.\ \ref{fig:res_vs_mu}(a)] indicates that, also in this case, $\mathrm{TOP}$ impurities dominate the transport.

If instead the sample has only critical $\mathrm{REC}$ impurities ($\tilde{\ee}_{\mathrm{REC}}\!=\!0$) [full black circles in Fig.\ \ref{fig:res_vs_mu}(b)], the same sharp resistivity peak appears at zero chemical potential, with a $\sim 95\%$ drop at $\mu\!=\!0$ when the $\mathrm{REC}$-impurity fraction is lowered to $5\%$, and the mixture contains $95\%$ of $\mathrm{VAC}$ impurities [empty black circles in Fig.\ \ref{fig:res_vs_mu}(b)]. However, as in the noninteracting case, $\mathrm{VAC}$ impurities do contribute to the resistivity for finite $\mu$. Indeed, from Fig.\ \ref{fig:res_vs_mu}(b) it is clear that the $\mathrm{VAC}$-impurity contribution dominates for $|\mu| > 3\,\mathrm{eV}$, owing to stronger Kondo scattering of graphene electrons with symmetric than with asymmetric vacancies. Remarkably, when the impurity energy is tuned to $\eeH^*$, a central Kondo--induced resistivity peak appears slightly away from charge neutrality [squares in Fig.\ \ref{fig:res_vs_mu}(b)]. This signature originates from Kondo scattering of graphene electrons by $\mathrm{REC}$ impurities, as can be inferred from the amplitude drop with increased symmetric--impurity fraction $n$ [Fig.\ \ref{fig:res_vs_mu}(b) inset].

This distinction between inversion--symmetric $\mathrm{HS}$ impurities and non--inversion--symmetric $\mathrm{VAC}$ impurities away from charge neutrality appears because, as shown in Fig.\ \ref{fig:thetasq}(a), the latter are decoupled only from zero--energy states at the Dirac points, due to $C_{3v}$ symmetry. As the chemical potential moves away from these points, those energy states become irrelevant for transport, and impurity scattering begins to dominate. $\mathrm{HS}$ impurities, on the other hand, are decoupled from entire branches across the Brillouin zone due to inversion symmetry [Figs.\ \ref{fig:thetasq}(c) and (d)], and thus from states of all energies. As a result, coherent transport remains the dominant mechanism for all chemical potentials within the low--energy approximation.

\section{Conclusions}\label{sec:conclusions}
We have studied the thermodynamic, spectral and linear transport signatures of dilute magnetic vacancies and adatoms in mesoscopic graphene. Our numerical results indicate that the quantum--critical\cite{fritz_vojta_kondo_in_graphene,fritz_vojta_prb_2004, fritz_vojta_prb_2_2004} behavior of magnetic impurities in neutral graphene is in direct correspondence with the single--particle picture of nonmagnetic impurities. Local interactions produce a level shift $|\ee_I^*|$, corresponding to the critical level energy, and renormalize the impurity-graphene hybridization. Quantum criticality is analogous to bound--state formation by resonant scatterers,\cite{ruiz-tijerina_2016} and in the case of top adsorbates and reconstructed vacancies introduces a sharp peak in the local density of states that scales with energy as $|\omega|^{-1}$. However, no Kondo physics is observed in the absence of a back gate,\cite{gonzalez-buxton_prb_1998} and the zero--energy resonance at criticality produces $\sim D^2T^{-2}$ resistivity scaling, leading to full insulating behavior at zero temperature, in stark contrast to the well--known Kondo resistivity plateau.

Away from charge neutrality, the system enters a Kondo phase that strongly depends on the impurity symmetry: While asymmetric impurities, such as top--adsorbates and reconstructed vacancies, remain in the Kondo phase for a wide range of back gate voltages, symmetric vacancies and hollow--site adsorbates exhibit Kondo correlations only within parameter--dependent limits. Nonetheless, symmetric and non--symmetric impurities display comparable, and experimentally accessible Kondo temperatures ranging from order $1\,\mathrm{K}$ to order $10\,\mathrm{K}$ for realistic parameters,\cite{wehling_prl_2010,yuan_prb_2010} defying earlier expectations.\cite{uchoa_prl_2011,po-wei-lo_prb_2014}

As in the case of nonmagnetic (noninteracting) impurities, symmetry is critical to the system electronic transport. Unreconstructed vacancies and other $C_{3v}$--symmetric impurities remain decoupled from graphene states at the $K$ and $K'$ points, which in charge neutrality remain available for electrons to move coherently through the sample. In the presence of a back gate, however, our results indicate that symmetric vacancies will contribute strongly to the resistivity through Kondo scattering. In contrast, inversion--symmetric impurities, such as hollow--site adsorbates, are decoupled from entire momentum branches across the Brillouin zone, and thus never contribute to the resistivity. Although this means that symmetric impurities cannot be probed in the bulk through resistivity measurements,\cite{ferreira_prb_2016} their remarkable properties in criticality and in the Kondo regime can be measured by means of local methods, such as scanning tunneling microscopy (STM).\cite{ren_nanolett_2014}

\begin{acknowledgments}
The authors thank Caio Lewenkopf for enlightening discussions. D.A.R.T.\  thanks S.\ E.\ Ulloa for fruitful discussions regarding the transport formalism, and acknowledges financial support by the Brazilian agency CAPES and by the Lloyd's Register Foundation Nanotechnology grant. L.G.G.V.D.S.\ acknowledges financial support by CNPq (grants No.\ 307107/2013-2 and 449148/2014-9), PRP-USP NAP-QNano and FAPESP (grant No.\ 2016/18495-4).
\end{acknowledgments}

\appendix

\section{Real--space impurity-graphene couplings }\label{app:impurities}
Here we present the expressions for the impurity-graphene couplings $\Hig$ in real space for the different impurity types ($I=\mathrm{TOP},\,\mathrm{HS}\,,\mathrm{VAC},\,\mathrm{REC}$). Without loss of generality we set the origin of our coordinate system at the impurity site. When the impurity sits at or on top of a lattice site ($\mathrm{TOP}$, $\mathrm{VAC}$ and $\mathrm{REC}$) we call the corresponding sublattice $A$.

$\mathrm{TOP}$ impurities couple to a single site as
\begin{equation}\label{eq:topA}
    \HtopA = V\sum_{s}\left\{d_{\mathrm{TOP},s}^\dagger a_s(0) + a_s^\dagger(0) d_{\mathrm{TOP},s} \right\}.
\end{equation}
A $\mathrm{VAC}$ impurity will couple identically to all three surrounding sublattice $B$ sites located at $a\uu_j$ as
\begin{equation}\label{eq:vacA}
    \HvacA = V\sum_{s}\sum_{j=1}^3d_{\mathrm{VAC},s}^\dagger b_s(a\uu_j) + \text{H.\ c.} 
\end{equation}
For the case of a non--symmetric $\mathrm{REC}$ impurity, we consider that the $sp^2$ orbital of the $B$--sublattice carbon atom at $a\hat{\vec{u}}_l$ will couple to the $\pi$ orbitals of the two $B$--sublattice carbons at $a\hat{\vec{u}}_j$ ($\l \ne j$) through
\begin{equation}\label{eq:recA}
    \HrecA(l) = V\sum_{s}\sum_{j\ne l}^3 d_{\mathrm{REC},s}^\dagger b_s(a\uu_j - a\uu_l) + \text{H.\ c.} 
\end{equation}
Finally, $H$ impurities couple identically to both sublattices:
\begin{equation}\label{eq:hs}
    \Hhs =V\sum_{s}\sum_{j=1}^3 d_{\mathrm{HS},s}^\dagger \big[a_s(a\uu_j) + b_s(-a\uu_j)\big] + \text{H.\ c.}
\end{equation}
In Fourier space we have
\begin{subequations}
\begin{equation}
    \HtopA = V\sum_{\vec{k},s}d_{\mathrm{TOP},s}^\dagger a_{\vec{k}s} + \text{H.\ c.} ,
\end{equation}
\begin{equation}
    \HvacA = V\sum_{\vec{k},s}\Phi(\vec{k})d_{\mathrm{VAC},s}^\dagger b_{\vec{k}s} + \text{H.\ c.} ,
\end{equation}
\begin{equation}
    \HrecA(l) = V\sum_{\vec{k},s}\exp{-i\vec{k}\cdot\uu_l}\left[\sum_{j\ne l}\exp{ia\vec{k}\cdot\uu_j}\right] d_{\mathrm{REC},s}^\dagger b_{\vec{k}s} + \text{H.\ c.} ,
\end{equation}
\begin{equation}
\begin{split}
    \Hhs =&V\sum_{s} d_{\mathrm{HS},s}^\dagger \big[\Phi(\vec{k})a_{\vec{k}s} + \Phi^*(\vec{k})b_{\vec{k}s}\big] + \text{H.\ c}.
\end{split}
\end{equation}
\end{subequations}
Applying the transformation (\ref{eq:unitary_2band}) we obtain Eqs.\ (\ref{eq:vVAC}), (\ref{eq:vREC}), (\ref{eq:vTOP}),  and (\ref{eq:vHS}), respectively.

\bibliography{bibliography}

\end{document}